\documentclass[aip,jap,reprint,twocolumn,letterpaper]{revtex4-1} % for review and submission  
\setcitestyle{square}
\usepackage{graphicx}  % needed for figures
\usepackage{float}
\usepackage{dcolumn}   % needed for some tables
\usepackage{bm}        % for math
\usepackage{amssymb}   % for math
\usepackage{amsmath}
\usepackage{isomath}
\usepackage[usenames, dvipsnames]{color}
\usepackage{xcolor}
\usepackage{soul}
\usepackage{physics}
\usepackage{textcomp}
\usepackage[normalem]{ulem}
\begin{document}
%\widetext
\title{Experimental demonstration of topological waveguiding \\
in elastic plates with local resonators}% Force line breaks with \\

\author{Rajesh Chaunsali}
\affiliation{${}^1$Aeronautics and Astronautics, University of Washington, Seattle, WA, USA, 98195-2400}
\author{Chun-Wei Chen}
\thanks{Equally contributed first author}
\affiliation{${}^1$Aeronautics and Astronautics, University of Washington, Seattle, WA, USA, 98195-2400}
\author{Jinkyu Yang}
\email{jkyang@aa.washington.edu}
\affiliation{${}^1$Aeronautics and Astronautics, University of Washington, Seattle, WA, USA, 98195-2400}
\date{\today}% It is always \today, today,

\begin{abstract}
It is recent that the emergence of topological insulators in condensed matter physics has inspired analogous wave phenomena in mechanical systems, mostly in the setting of discrete lattice models. Here we report a numerical and experimental demonstration of topological waveguiding in a continuum plate. We take a ubiquitous design of a bolted elastic plate and show that such a design allows us to invoke the pseudo-spin Hall effect at remarkably low frequencies. We harness the complex interaction of the bolts and the plate to show the existence of a pair of double Dirac cones near the resonant frequency of the bolt. The manipulation of bolted patterns results in the opening of multiple topological bandgaps, including a complete bandgap that forbids all plate modes. We demonstrate that inside this bandgap, the interface between two topologically distinct zones can guide flexural waves crisply around sharp bends.

\end{abstract}
%\pacs{45.70.-n 05.45.-a 46.40.Cd}% PACS, the Physics and Astronomy
\maketitle

\section*{Introduction}
The discovery of topological insulators in condensed matter physics has prompted a new notion of topology in association with the intrinsic dispersion behavior of a material~\citep{Hasan2010, Qi2011}. By using this concept, one can characterize the dispersion behavior of an infinite (i.e., ``bulk'') material, which consequently provides a tool to predict the response at the ``boundaries'' of a finite material. This ``bulk-boundary correspondence''  leads to a topologically protected boundary response of a non-trivial bulk, thereby offering a degree of robustness. At a physical level, the topological insulator has thus shown an exotic state, in which a robust and directional current flows along the material's boundary, while it is forbidden in the bulk.

This tool of topology has paved a way for researchers to control the flow of energy in other areas, such as photonics \citep{Lu2014} and acoustics \citep{Yang2015, He2016, Fleury2016, Zhang2016}. It has also given an impetus to a new way of designing elastic systems \citep{Nash2015, WangP2015, WangY2015, Chaunsali2016, Huber2015, Pal2016, Kariyado2016, Chaunsali2017, Vila2017, Wu2018, Liu2018, Zheng2018}. These topological structures---mostly in the setting of discrete lattices---aim at manipulating elastic vibrations and offer a tremendous degree of flexibility in controlling their dynamic responses. Therefore, these prototypical systems are excellent candidates for tabletop designs, in which topological physics can be systematically investigated \citep{Huber2016}.
%At the same time, wave manipulation capabilities of these elastic systems promise potential future applications in energy harvesting, vibration isolation, and structure health monitoring \citep{Huber2016}. \textbf{This last sentence is somewhat contradicting to our claim on the practicality of our plate model that follows next.}
However, one of the outstanding challenges in the topological manipulation of elastic systems is to have control over several types of wave modes that can exist in elastic solids. This becomes even more relevant when one goes beyond the discrete lattice models and considers continuum structures such as plates. Generally, there are two ways to deal with this. One is to capitalize this complexity to invoke topological effects by judiciously combining different wave modes \citep{Mousavi2015, Miniaci2017}. The other is to neglect other wave modes and to focus only on a ``decoupled'' mode, e.g., the flexural mode in thin plates \citep{Torrent2013, Pal2017, Chaunsali2018, Jin2018} and the symmetric modes in thick plates \citep{Brendel2017}. While the former approach is more suitable for high operating frequencies, the later is prone to energy leakage due to the discounted wave modes.   

\begin{figure*}[t]
\centering
\includegraphics[width=0.9\linewidth]{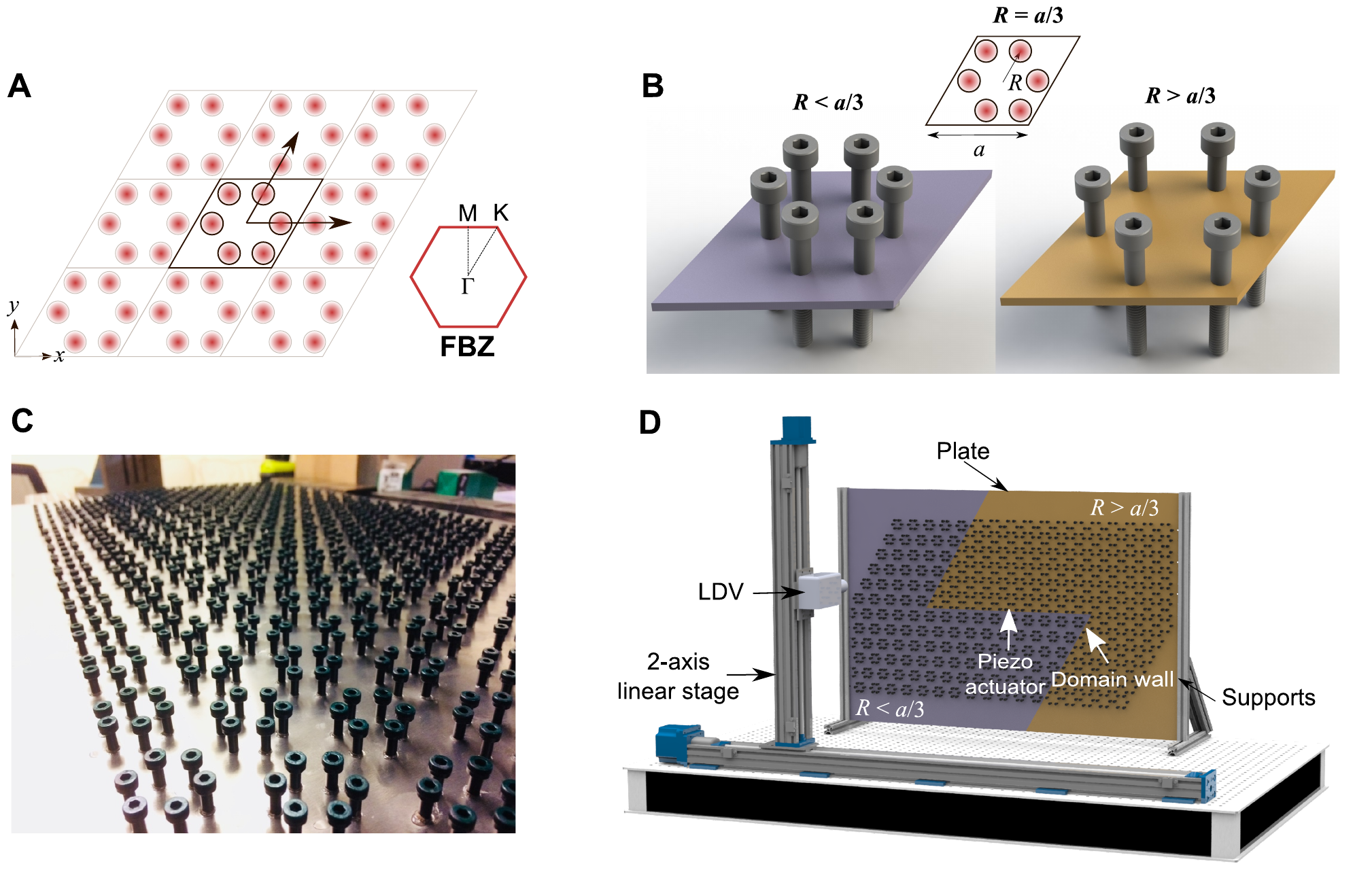}
\caption{System configuration. \textbf{A}, A hexagonal lattice arrangement with the unit cell containing six elements. The corresponding first Brillouin zone (FBZ) is indicated. \textbf{B}, This unit cell is perturbed around $R=a/3$ to obtain two topologically distinct unit cells. Bolts act as local resonators mounted on a thin plate. \textbf{C}, A snapshot of the actual system with multiple bolts mounted on the plate. \textbf{D}, The experimental setup, in which a domain wall is created between two topologically distinct lattice patterns in yellow and violet. The plate is excited by a piezo actuator positioned at the center, and out-of-plane velocity is measured point-by-point by an LDV mounted on a 2-axis linear stage.}
\label{fig1}
\end{figure*} 

A natural question is whether there exists a way to manipulate elastic waves at low frequencies yet avoiding the wave leakage into other modes. This can be of significant importance, because low-frequency flexural modes in plates typically exhibit large amplitude, and their efficient manipulation can be useful for several engineering applications, such as energy harvesting, vibration isolation, and structure health monitoring. However, lowering the operating frequencies generally demands large lattice sizes due to the Bragg condition. This can pose challenges especially under stringent size limitations of the wave medium. Therefore, the efficient topological manipulation of low-frequency waves in elastic systems remains a formidable challenge to date.  

This quest hints us to leverage a locally resonant waveguide, in which the bandgaps created by the locally resonant elements  can result in filtering of selected wave modes at low frequencies \citep{Liu2000}. In this study, we complement the topological physics by the mechanism of local resonance to demonstrate the topological waveguiding of low-frequency waves. Recent numerical studies have shown that \textit{out-of-plane} coupling of resonators and a thin plate can be used to realize Dirac cones and invoke associated topological physics for flexural waves \citep{Torrent2013, Pal2017, Chaunsali2018}. However, \textit{in-plane} modes have been often neglected in these studies, despite the fact that their presence can cause leakage of flexural waves. While recent numerical studies provide novel solutions to this issue \citep{Chen2017, Foehr2018}, the experimental realization of such locally resonant topological systems has not been yet achieved. 

Here we take a ubiquitous design of bolted plates to tackle the aforementioned challenges. The design is rich in physics as it not only allows the out-of-plane coupling between the mounted bolts and the plate, but also includes the effect of their bending on in-plane modes. We employ the zone-folding technique \citep{Wu2015} to invoke the $C_6$ symmetry-protected pseudo-spin Hall effect and manipulate flexural waves by strategically arranging the bolts on the plate. Remarkably, this bolted plate system forms a pair of \textit{double} Dirac cones \citep{Sakoda2012} in the low-frequency range of the bolt's bending resonance. Such double Dirac cones are a key ingredient in the creation of two pseudo-spins propagating in opposite directions in the topological waveguide. 
We numerically investigate the opening of multiple topological bandgaps by manipulating the bolt arrangement on the plate, one of which is notably a complete bandgap that prevents the leakage of flexural waves into other modes. We experimentally verify efficient waveguiding capability of this bolted plate along a path with sharp bends, when excited at frequencies inside these topological bandgaps.
 
\section*{Results}
\textbf{Design of the locally resonant plate with a topological waveguide.} Our system consists of a thin plate and bolts that are tightly fastened to form a periodic pattern (see Methods for details). As a starting point, we take a hexagonal lattice pattern shown in Fig.~\ref{fig1}A. We focus on a large unit-cell representation, i.e., a rhombus-shaped unit cell (lattice constant $a$) that contains six bolts to facilitate zone-folding. We then perturb this unit cell by varying the circumferential radius $R$ of the mounted bolts around $R=a/3$ (Fig.~\ref{fig1}B). This is to obtain two topologically distinct unit-cell configurations of the bolted plate. Note that, in the process, we keep the $C_6$ symmetry intact in the unit cell, so that we obtain degenerate modes in the dispersion to be shown later. These two types of unit cells, specifically $R=0.8a/3$  and $1.1a/3$  in this study, are then tessellated to form a periodic pattern in the plate (Fig.~\ref{fig1}C). We construct a domain wall consisting of three linear segments (one horizontal and two inclined at 60$^\circ$) by joining these two types of periodic patterns (Fig.~\ref{fig1}D). The plate is excited with a piezo actuator attached in the middle of the horizontal domain wall, and a laser Doppler vibrometer (LDV) scans the plate's vibrations via a two-axis linear stage.  

\begin{figure*}[t!]
\centering
\includegraphics[width=0.9\linewidth]{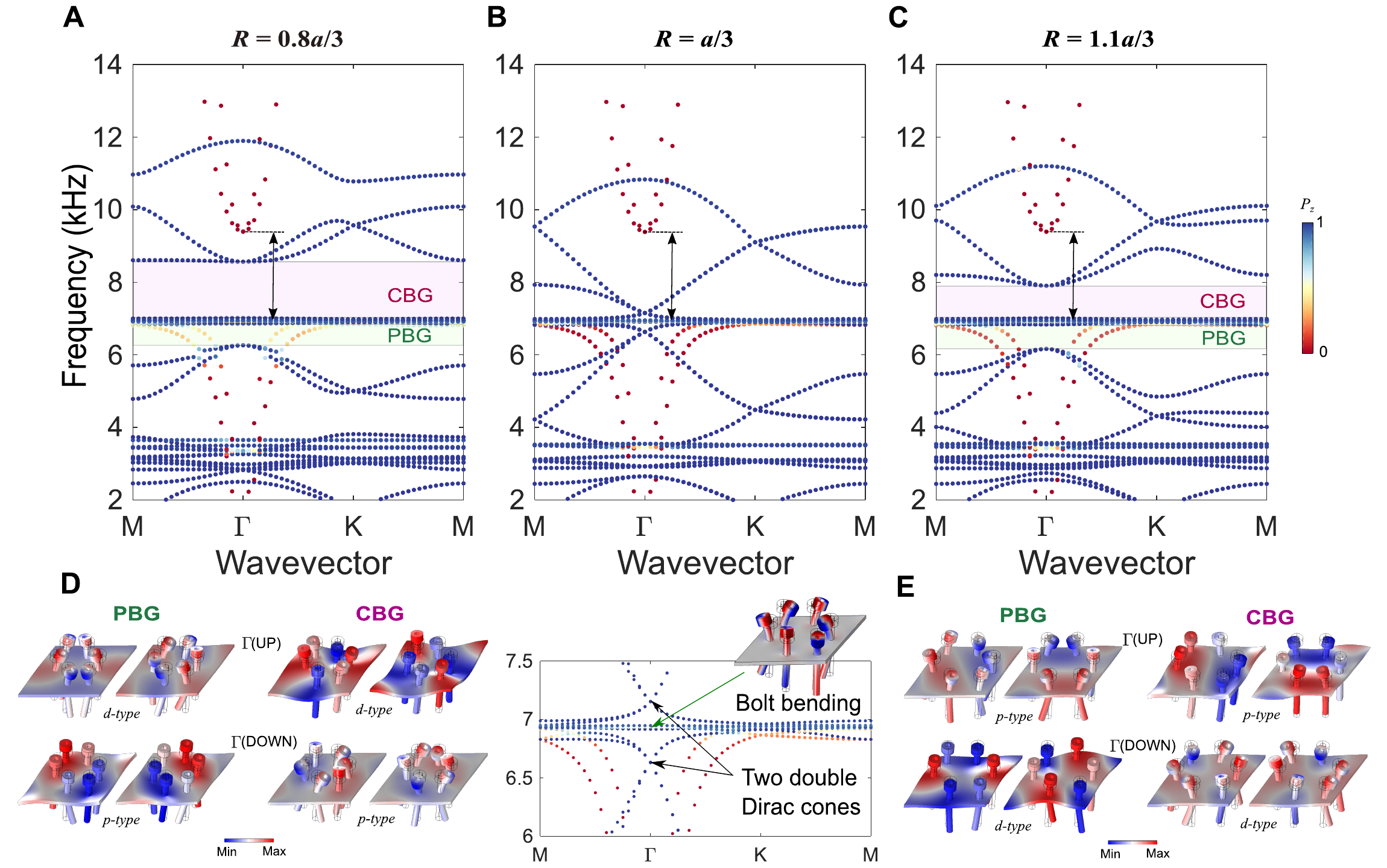}
\caption{Presence of multiple bandgaps and their inversion in the unit-cell dispersion. \textbf{A}, A case with $R<a/3$ leading to the emergence of two bandgaps. Color of the markers indicates the polarization of plate mode, i.e., it is nearly 1 for out-of-plane and nearly 0 for in-plane plate modes. The lower bandgap is a partial bandgap (PBG) due to the presence of other nearly in-plane plate modes. The upper bandgap is a complete bandgap (CBG) for all plate modes. The two-sided arrow indicates the locally resonant bandgap for in-plane modes. \textbf{B}, In the hexagonal lattice arrangement, i.e., $R=a/3$, these bandgaps close and form two distinct double Dirac cones. Zoomed-in view is in the inset below to show the bolt bending mode between two double Dirac cones. \textbf{C}, When $R>a/3$, the two double Dirac cones open again to form two distinct bandgaps. \textbf{D}, The mode shapes (out-of-plane) at the $\mathrm{\Gamma}$ point for the bandgaps shown in \textbf{A}. $\mathrm{\Gamma}$(UP) and $\mathrm{\Gamma}$(DOWN) indicate the upper and the lower edges of the bandgap, respectively. Note that two degenerate $p$-type modes have smaller frequency than two degenerate $d$-type modes for both bandgaps. \textbf{E}, The mode shapes at the $\mathrm{\Gamma}$ point for the bandgaps shown in \textbf{C}. These mode shapes are inverted with respect to the ones shown in \textbf{D}.}
\label{fig2}
\end{figure*} 

\textbf{Unit-cell dispersion and band inversion.} In Fig.~\ref{fig2}, we numerically show dispersion diagrams for the unit-cell configurations with radii varying across $R=a/3$. For $R=0.8 a/3$, we observe two distinct bandgaps in the range of 6.2 -- 8.5 kHz in Fig.~\ref{fig2}A. The colormap, indicating the polarization of plate modes, confirms that the bandgaps are for out-of-plane modes (blue markers). See Methods for the calculation of the polarization index. The lower bandgap is a partial bandgap (PBG) because it supports nearly in-plane modes ($S_0$ and $SH_0$) shown with red and yellow markers. Notably, the upper one is a complete bandgap (CBG). This is because the bending-dominated local resonance of the bolts occurs near 6.9 kHz (see the horizontal blue markers in Fig.~\ref{fig2}A), and it is coupled with in-plane modes of the plate (further explanations to follow next). This results in a locally resonant bandgap (see the two-sided arrow) for ``undesired'' in-plane modes (red and yellow markers) in the range of 6.9 -- 9.4 kHz. Consequently, we observe a complete bandgap between 6.9 kHz to 8.5 kHz for all plate modes. 

This is an example of judicious engineering of overlapping two types of bandgaps: locally resonant bandgap for in-plane modes and Bragg bandgap for out-of-plane modes in the same system. We call the later Bragg bandgaps because those emerge solely when the translational symmetry in the system is changed by varying the radius across $R=a/3$. This point can be further justified when we plot the dispersion diagram for the case of the perfectly hexagonal system, i.e., with $R=a/3$, in Fig.~\ref{fig2}B. We observe that both bandgaps (denoted by PBG and CBG above) close.  However, the bandgap for in-plane modes (red markers) remains intact, because it is a locally resonant bandgap caused by the local bending of the bolts, which we do not change in the process (see the two-sided arrow). Remarkably, near the resonant frequency of the bolts, we observe the formation of a pair of distinct double Dirac cones (see the panel below Fig.~\ref{fig2}B for the zoomed-in view of the dispersion curves along with the mode shape of the resonant bolts). To the best of the authors' knowledge, creation of such multiple double Dirac cones at different frequencies in an elastic system has not been reported in the literature. %Observation of this accidental phenomena in a ubiquitous design of a bolted plate makes it even more striking. 

\begin{figure*}[t!]
\centering
\includegraphics[width=0.9\linewidth]{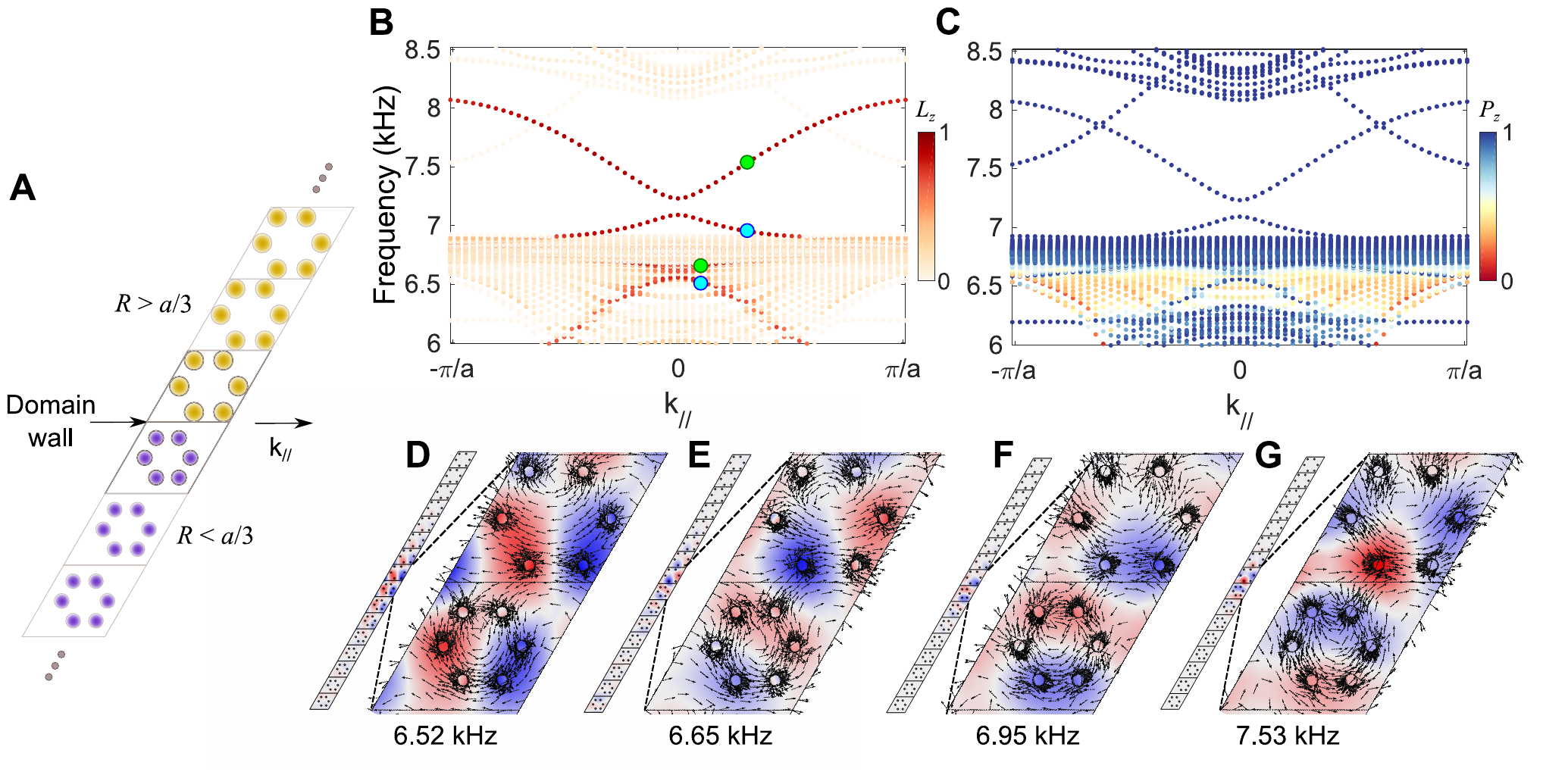}
\caption{Emergence of topological modes at the domain wall. \textbf{A}, A supercell made by placing two topologically distinct lattices adjacently at the domain wall. \textbf{B}, Eigenfrequencies of the supercell vs. $k_{//}$, the wavevector parallel to the domain wall. The colormap indicates the localization index ($L_z$). Higher the value it has, the more localized are the out-of-plane modes at the domain wall. The upper region (i.e., CBG) shows clear localized modes. The lower region (PBG) also shows localized modes, but those are mixed with other extended modes. \textbf{C}, The same eigenfrequencies diagram, but with a different colormap to show the polarization of plate modes. It denotes 1 (0) for out-of-plane (in-plane) plate modes. \textbf{D-G}, Localized out-of-plane mode shapes at some selected frequencies inside PBG and CBG (marked by circles in \textbf{B}). The arrows indicate the time-averaged mechanical energy flux to confirm their spin nature.}
\label{fig3}
\end{figure*}

In Fig.~\ref{fig2}C, we plot the dispersion diagram for the case when we further increase the unit-cell radius to $R=1.1a/3$. We choose this radius to avoid the bolts approaching too close to the unit cell boundary, and at the same time to ensure an overlap with the bandgaps established in $R=0.8a/3$ (compare the locations of bandgaps between Figs.~\ref{fig2}A and C). We observe that the two double Dirac cones open again and form two bandgaps. However, these are different from the ones observed in Fig.~\ref{fig2}A in terms of topology. For further investigation, we revisit Fig.~\ref{fig2}A, examine the out-of-plane mode shapes, and plot them in Fig.~\ref{fig2}D. For the lower bandgap, i.e., PBG, we extract the mode shapes of its edges at the $\mathrm{\Gamma}$ point. We observe that two $p$-type modes are degenerate at the lower edge, i.e., $\mathrm{\Gamma}$(DOWN), and two $d$-type modes are degenerate at the upper edge, i.e., $\mathrm{\Gamma}$(UP). Naturally, the $d$-type modes show prominent bending of the bolts, approaching the vicinity of local resonance region around 6.9 kHz. A similar pattern is observed for the upper bandgap (CBG) as well, wherein $d$-type modes are at the upper edge. Sandwiched between these two regions are the locally resonant bending modes of the bolts.

In Fig.~\ref{fig2}E, we plot the mode shapes at the aforementioned band edges at the $\mathrm{\Gamma}$ points, but for the configuration with $R=1.1a/3$ shown in Fig.~\ref{fig2}C. A similar degeneracy of modes is observed. However, $p$-type modes are at higher frequencies than $d$-type modes for both bandgaps. This so-called band-inversion is a crucial ingredient of the topological phenomena at work and makes this configuration topologically non-trivial as per the analogy to atomic orbitals where $p$-type orbitals are at lower energy (frequency) than $d$-type orbitals in general. This whole mechanism is therefore reminiscent of the pseudo-spin Hall effect proposed in Ref.\citep{Wu2015}, where the degeneracy of $p$- and $d$-type modes is used for creating two counter-propagating pseudo-spin modes at the domain wall between two topologically distinct unit cells. We also emphasize that these topological bandgaps are formed in a continuum plate at a low frequency regime, i.e., in the vicinity of the bending mode of the bolts.

%, we would like to appreciate this topological phenomena, which occurs in a continuum plate at low frequency regimes, i.e., in the vicinity of the low order bending mode of the bolts.

%1. We also confirm that this topological waveguiding is achieved in a continuum plate at a remarkably low frequency regime, which is governed mainly by the bending mode of the bolts and thus, can be further adjusted by changing bolt mass and geometry.

%2. At this juncture, we would like to appreciate this topological phenomena, which occurs in a continuum plate at low frequency regimes, i.e., in the vicinity of the low order bending mode of the bolts.

\begin{figure*}[t!]
\centering
\includegraphics[width=0.9\linewidth]{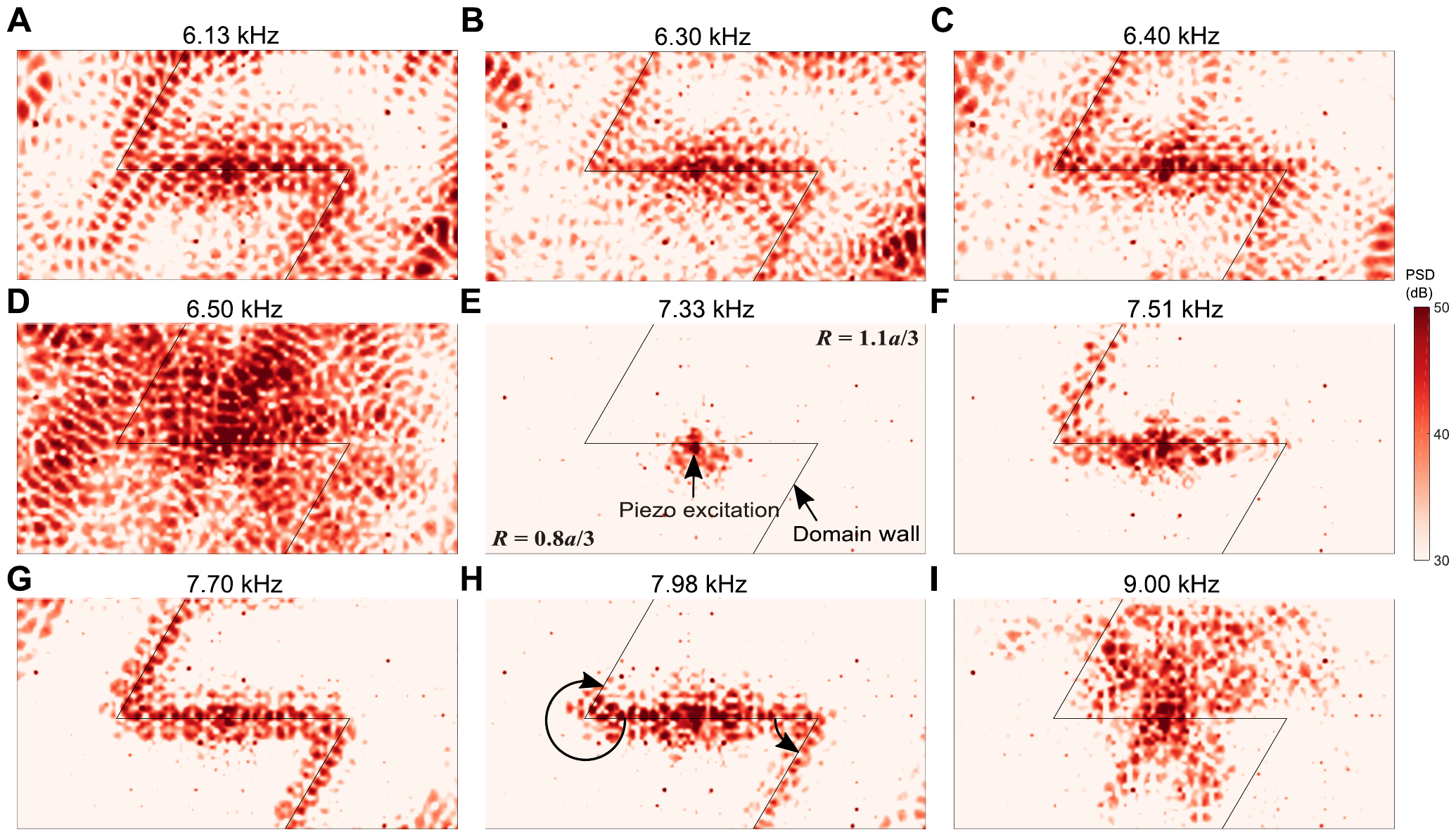}
\caption{Experimentally obtained steady-state transmission maps at multiple excitation frequencies of the piezo actuator located at the domain wall. Colors indicate the power spectral density (PSD) calculated from the measured out-of-plane velocity of the plate. \textbf{A-C}, Wave localization at the domain wall for the frequencies inside PBG. \textbf{D-E}, Disappearance of this localization indicates the dispersion region between PBG and CBG, where no localized topological mode exists at the domain wall. \textbf{F-H} Emergence of wave localization at the domain wall once again indicates the region of CBG. \textbf{I}, Wave transmission at a frequency above CBG.}
\label{fig4}
\end{figure*} 

\textbf{Emergence of topological interface state.} Now moving to the multicell configuration, we first take a supercell consisting of eight unit cells of each type and join them to form a domain wall as shown in Fig.~\ref{fig3}A. We apply a periodic boundary condition in parallel to the domain wall to investigate the wave dynamics of this tessellated system. Because we have two topologically distinct lattices, we expect a localized plate mode at the domain wall as per the bulk-boundary correspondence in topological systems \citep{Hasan2010}. In Fig.~\ref{fig3}B, we show the eigenfrequencies of the supercell as a function of wavevector $k_{//}$. The high value of the localization index (see Methods for the calculation) in the colormap clearly highlights the presence of localized modes at the domain wall (red markers) inside both bandgaps (PBG and CBG). The modes in the upper region are desolate as expected in the complete bandgap. However, the localized modes in the lower region are populated with other modes in this partial bandgap. 

To verify the polarization of these localized modes, we replot Fig.~\ref{fig3}B in Fig.~\ref{fig3}C but with a different colormap indicating the wave polarization, as was used in Fig.~\ref{fig2}. We clearly observe that the localized modes at the domain wall are out-of-plane modes (blue markers).  Again, the upper region shows clean out-of-plane modes. The additional branch in this bandgap corresponds to the modes at the other ends of the supercell, and therefore not of any important considerations here. The lower region, however, shows weaker out-of-plane localized modes, and those are intermixed with other in-plane modes (red and yellow markers). This fact leads us to understand the difference of topological wave localization between PBG and CBG. The advantage of having the CBG is evident as it facilitates a clean wave localization at the domain wall and eliminates the possibility of wave leakage into other modes. Note in passing that there is a small gap at the $k_{//}=0$ point in both regions, where there is no localization at the domain wall. As highlighted in the previous studies \citep{Wu2015, Chaunsali2018}, this is due the breakage of crystalline $C_6$ symmetry at the domain wall. By employing engineering tricks, such as a slowly-varying (i.e., gradient) domain wall, one can also reduce this small gap at the $k_{//}=0$ point and impart greater protection to the topological modes.

We proceed to investigate the mode shapes of the aforementioned localized modes. We plot out-of-plane displacements at selected frequencies marked as circles in Fig.~\ref{fig3}B. Figures~\ref{fig3}D-E correspond to the frequencies inside PBG, and Figs.~\ref{fig3}F-G correspond to the frequencies inside CBG, all arranged in the order of increasing frequencies. It is clear that these modes are localized at the domain wall. Insets show zoomed-in views of the mode shapes on which the superimposed arrows indicate in-plane time-averaged mechanical energy flux ($I_j=-\sigma_{ij}v_j$, where $\sigma_{ij}$ and $v_j$ are stress tensor and velocity vector, respectively) over a harmonic cycle. It is evident that these modes have spin characteristics. More specifically, in Figs.~\ref{fig3}D,F that correspond to the cyan markers in Fig.~\ref{fig3}B, we observe clockwise spins. In Figs.~\ref{fig3}E,G (corresponding to the green markers in Fig.~\ref{fig3}B), we witness counterclockwise spins (see the lattices with smaller $R$ for clearer visualization). As the time-reversal symmetry is intact in our system, it is obvious that we will have exactly the opposite spins at these frequencies for a negative wavevector $k_{//}$. Therefore, judging from the group velocity, i.e., the slope of the dispersion branch at the green circles in Fig.~\ref{fig3}B, we expect that the counterclockwise spin propagates along the domain wall towards positive wavevector, while the clockwise spin propagates in the opposite direction. We will look for these spins in the experimental results later.

\begin{figure*}[t!]
\centering
\includegraphics[width=0.9\linewidth]{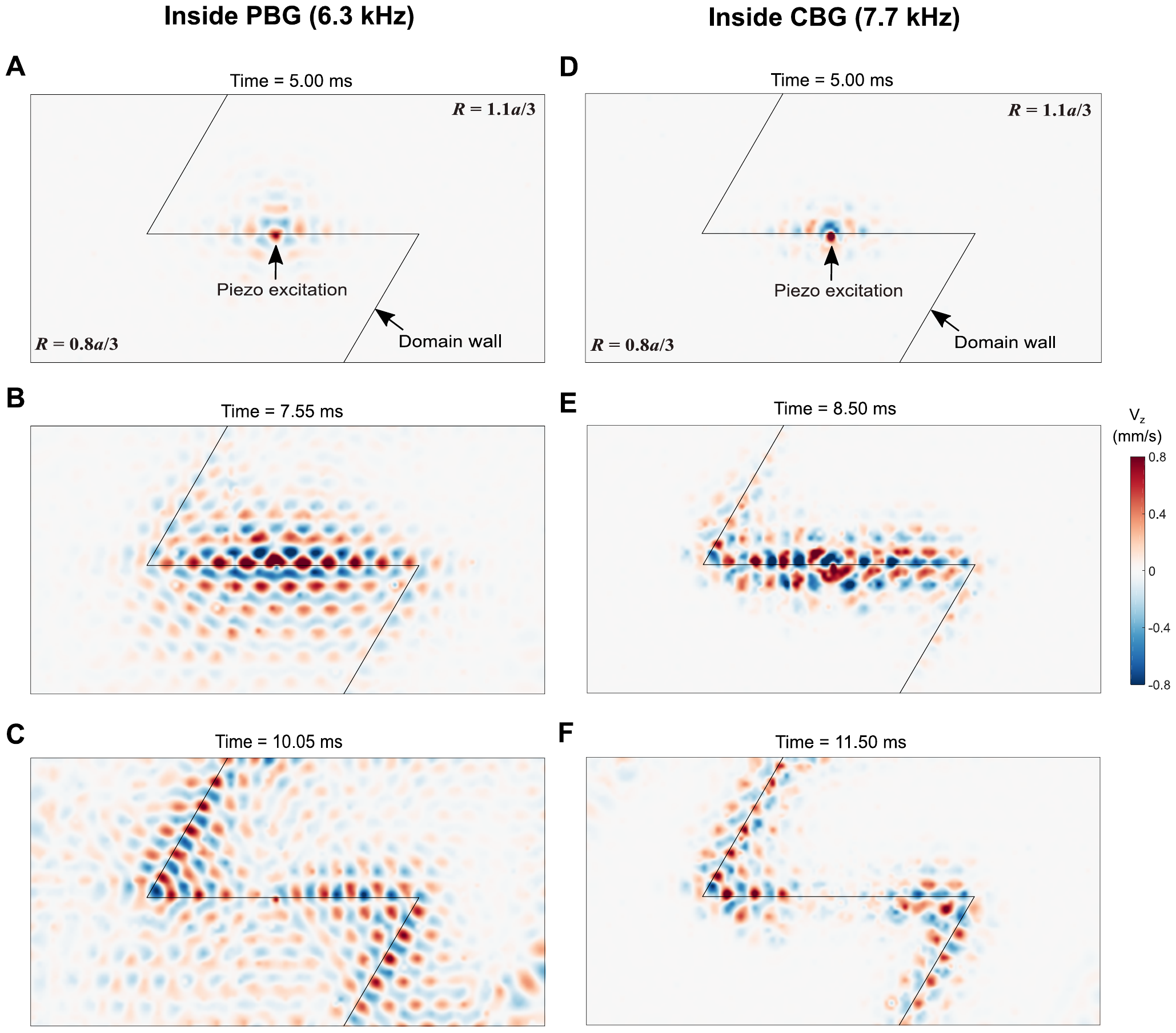}
\caption{Experimentally obtained transient wavefields. \textbf{A-C}, Velocity wavefields when a Gaussian input centered at 6.3 kHz is injected by the piezo actuator in the middle of the domain wall. \textbf{D-F}, Wavefields for the Gaussian input centered at 7.7 kHz.}
\label{fig5}
\end{figure*} 

\textbf{Steady-state transmission.} Building from this numerical result, we experimentally verify the existence of topological modes at the domain wall for both PBG and CBG regions. In Fig.~\ref{fig4}, we show the evolution of steady-state transmission plots on the plate, as we inject vibrational energy through a piezoelectric actuator attached at the center of the domain wall and increase its input frequency (see Methods for experimental details). At low excitation frequencies (Figs.~\ref{fig4}A-C), we observe that energy is primarily localized at the domain wall.  %This localization spans all along the domain wall indicating a decent wave transmission through sharp bends. However, note that the transmission map changes with the input frequency. In Fig.~\ref{fig4}D, which is for a higher frequency, we observe that this wave localization is lost and wave propagates all around the plate. As the input frequency gets higher, a drastic attenuation in wave transmission is observed in Fig.~\ref{fig4}E.  Remarkably, for the frequencies above this point we again observe wave localization along the domain wall (Figs.~\ref{fig4}F-H). The wave localization is again lost and we observe transmission all along the plate at 9 kHz as shown in Fig.~\ref{fig4}I.
%This transmission trend is in good agreement with the wave localization phenomena discussed in Fig.~\ref{fig3}. 
This waveguiding pattern detected in this low frequency range corresponds to the PBG (see Figs.~\ref{fig3}B-C). We observe some traces of elastic energy far away from the domain wall. This can be attributed to the nature of this PBG as it allows other wave modes to co-exist in the system. %\textbf{These modes might be excited from non-ideal initial excitation or mode-conversion at boundaries.} 
In addition, the smaller size of this bandgap would generally mean higher localization length of the modes at the domain wall, and hence, the flexural wave would penetrate more into the bulk. Beyond this PBG, the wave localization is lost, marking the presence of global modes (extended in plate as shown in Fig.~\ref{fig3}D). At 7.33 kHz (Fig.~\ref{fig3}E), the wave is highly attenuated around the point of excitation.

Remarkably, for the frequencies above this point, we again observe wave localization along the domain wall (Figs.~\ref{fig4}F-H). This corresponds to the localization region in the CBG as indicated in Figs.~\ref{fig3}B-C. We clearly observe a persistent wave localization along the horizontal domain wall for a wide range of frequencies. Among these, the transmission at 7.70 kHz (Fig.~\ref{fig4}G) is the most robust, in which the wave follows the entire domain wall without any significant back-scattering around the sharp bends. 

It is worth pointing out that at 7.51 and 7.98 kHz (Figs.~\ref{fig4}F and H), the wave back-scatters around the sharp bends. A similar behavior was also observed recently in experiments on a mechanical counterpart of quantum valley Hall effect \citep{Qian2018}, where guided topological waves are back-scattered around a sharp bend for some frequencies inside the bandgap. However, here we observe that the wave is able to pass through the left bend at a lower frequency (Fig.~\ref{fig4}F) and through the right bend at a higher frequency (Fig.~\ref{fig4}H). This implies the preference of the topological wave propagation around one bend over the other, depending on the excitation frequency in the CBG. We explain this mechanism by the asymmetry of the two bends experienced by the spin waves. That is, as marked by the circular arrows in Fig.~\ref{fig4}H, the bending angle experienced by the counterclockwise spin that propagates towards the right is different from that by the clockwise spin towards the left. This explanation can be complemented by the following argument based on topology. In this symmetry-protected topological system, we have intentionally broken $C_6$ symmetry at the domain wall, including the sharp bends. This makes the modes less ``protected'' (i.e., less robust), thereby resulting in the spin-dependent transmission efficiency around the bends (see SM for further experimental results). 

%Nevertheless, this interesting and rich phenomena demands further future studies. For now, these observations lead us to conclude that a wide bandwidth of topological modes exists at the domain wall, but not all are robust around these sharp bends. 
% A similar behavior was also observed recently in experiments done on a mechanical counterpart of quantum valley Hall effect \citep{Qian2018}, where guided topological waves are back-scattered around a sharp bend for some frequencies inside the bandgap. \textbf{Supplementary document available?}
%This spin-dependent transmission efficiency around the bends can also be further explained by the following argument based on topology. 

%Overall, the transmission trend in Fig.~\ref{fig4} is in remarkable agreement with the wave localization phenomena predicted from the supercell analysis (Fig.~\ref{fig3}). Note that the subtle mismatch in the frequency regions between the numerical method and the experimental observation can be possibly due to the complexity of modeling the threaded contact between the bolts and plate (see supplementary details). Nonetheless, this simple model captures the effective contact stiffness reasonably well, thus elucidating the nature of bandgaps and topological localization in our bolted plate system. \textbf{This whole paragraph can be removed for the brevity of our manuscript.}   

\textbf{Robust waveguiding.} Now we demonstrate robust waveguiding capabilities of our system by capturing the transient wavefield when excited at the frequencies inside PBG and CBG. Navigating the steady-state profiles observed in the previous section, we choose two central frequencies of excitation, 6.3 kHz (in PBG) and 7.7 kHz (in CBG), and send Gaussian pulses from the piezo-actuator (see Methods for details). In Figs.~\ref{fig5}A-C, we show transient wave propagation on the plate at three different time steps inside PBG. We observe that the wave, emanating from the point of excitation at the middle of the domain wall, is effectively guided through the domain wall. Here, we verify that a counterclockwise spin propagates towards the right and a clockwise spin propagates towards the left (see Supplementary Movie 1). Note Fig.~\ref{fig5}C that there is some visible leakage into the bulk, complying well with the result from the steady-state response (Fig.~\ref{fig4}B).  In Figs.~\ref{fig5}D-F, we show transient wave propagation inside CBG. It is evident that this wave is also being guided along the domain wall without any significant back-scattering at the sharp bends (see Supplementary Movie 2). However, the key difference is that there is no evident leakage into the bulk due to the presence of CBG (Fig.~\ref{fig5}F). This result is consistent with the supercell analysis (as shown in Fig.~\ref{fig3}B) and the steady state response (Fig.~\ref{fig4}G).

%We also verify the spin-dependent transmission efficiency in the system by exciting the plate near 7.98 kHz (Fig.~\ref{fig4}H). We observe a clear back-scattering at the left bend (see Supplementary Movie 3), whereas the wave is able to pass through the right bend.

\section*{Discussion}
In this study we have demonstrated numerically and experimentally the capability of topological waveguiding in a plate with local resonators. By incorporating the local resonance effects in the design, we observe some novel phenomena in this study. First, low-frequency bending mode of the bolt leads to the creation of a pair of double Dirac cones, thus invoking topological effects at low-frequencies. Second, it enables us to deal with undesired in-plane plate modes, which makes it possible to clearly guide flexural waves along a topological waveguide, eliminating their leakage to in-plane modes. 
This is highly desirable in designing efficient waveguides for low-frequency flexural waves. In addition, we have also observed spin-dependent transmission efficiency of our topological waveguide around sharp corners for certain frequencies inside topological bandgaps. This is an interesting feature that can be exploited in applications, but at the same time, demands further exploration.

This simple and tunable design paves the way for a number of studies in future, especially with regard to studying the effect of disorder in such symmetry-protected topological systems. For example, the height of the bolts can be changed, or a nut can be attached to it, in order to introduce disorder in the system (see SM for a few examples). It is remarkable that even though a degree of disorder exists in the current setup in terms of bolt torques and plate threads, which in turn can change the effective coupling, we are still able to guide stress waves robustly. This is a hallmark of topological mechanical systems. 

\section*{Methods}
\textbf{Numerical simulations.}
We use commercial finite element software (COMSOL Multiphysics) to perform numerical simulations. We take nominal material properties of aluminum plate (Young's modulus $E=68.90$ GPa, density $\rho=2,700$ kg/m$^3$, Poisson ratio $\nu=0.33$) and steel bolts (Young's modulus $E=210.60$ GPa, density $\rho=7,800$ kg/m$^3$, Poisson ratio $\nu=0.30$). The threaded contact between the bolt and plate is modeled using a simple approach, which captures the realistic contact stiffness by modeling an effecting contact area (see SM for details). We mesh the model with tetrahedral elements. To obtain the dispersion diagrams in Fig.~\ref{fig2}A-C, we take a unit-cell and apply Bloch-Floquet boundary conditions and solve for eigenfrequencies along the irreducible Brillouin zone. % (M-$\mathrm{\Gamma}$-K-M). 
In the supercell analysis shown in Fig.~\ref{fig3}, we apply a Bloch-Floquet boundary condition along $k_{//}$ and keep the other boundaries free. 

To get more insights from the dispersion curves, we define two indices. First, the polarization index $P_{z}$ quantifies the out-of-plane plate modes. Second, the localization index $L_{z}$ quantifies the extent of localization of out-of-plane plate modes near the domain wall. The polarization index is defined as $P_{z}=\dfrac{\int_{V_u}{|u_{z}|^2 dV}}{\int_{V_{u}}(|u_{x}|^2+|u_{y}|^2+|u_{z}|^2)dV}$, where $V_{u}$ is the plate volume in the unit cell, and $u_{x}$, $u_{y}$, and $u_{z}$ are the displacement components in $x, y$, and $z$ axis, respectively. $P_{z}$ would be close to 1 for out-of-plane plate modes and near 0 for in-plane plate modes. 
The localization index is defined as $L_{z}=\dfrac{\int_{V_{s}*}{|u_{z}|^2 dV}}{\int_{V_{s}} |u_{z}|^2 dV}$, where $V_{s}$ is the entire plate volume of the supercell and $V_{s*}$ is the plate volume of two unit cells touching the domain wall (see the enclosed cells in Fig.~\ref{fig3}A with a darker solid line). It is evident that the higher value of $L_z$ implies the more localization of out-of-plane plate modes at the domain wall.

\textbf{Experimental fabrication and setup.}
The substrate is an aluminum 6061-T6 plate (91.4 $\times$ 61.0 cm and thickness $h=2$ mm). The rhombus-shaped unit cell has a dimension of $a=45$ mm. We drill six hexagonally arranged holes per unit cell using a CNC milling machine. We vary the spacing of hexagonally arranged bolts by using two circumferential radii $R=0.8a/3$ and $1.1a/3$. We then tap the holes manually and mount M4-0.7 mm black-oxide alloy steel socket head screws (91290A180, McMASTER-CARR) as the resonators. To facilitate the firm contact between the bolts and the threads of the plate, we add a drop of glue (Henkel Loctite adhesive) in the process of mounting bolts using a screwdriver (Supplementary Movie 4). 
% The resonator neck are $h_{neck}=30$ mm and $d_{neck}=4$ mm, whereas the resonator head are $h_{head}=4$ mm and $d_{head}=7$ mm.(20 mm of $h_{neck}$  is partially threaded. %
The partial thread on the screws comes handy, as it allows tightening them firmly on the plate without using any additional nuts. The total number of bolts mounted is 1320. We bond a piezo ceramic disc (SMD10T1F199R111, STEMiNC, diameter $10$ mm, and thickness $1$ mm) with silver epoxy adhesive on the plate as shown in Fig.~\ref{fig1}D. We mount the LDV (Polytec OFV 534) on the 2-axis linear stages and control its motion using MATLAB to measure vibrations on the plate. 

\textbf{Experimental measurements and processing.}
The LDV takes point-by-point measurements by moving in a square grid of 7.5$\times$7.5 mm. We measure 114 points horizontally and 60 points vertically, i.e., total of 6840 points. All measurements are synchronized with respect to the onset of the input voltage signal from the function generator, which sends a signal to the piezo actuator via an amplifier. For the steady-state transmission profiles shown in Fig.~\ref{fig4}, we send a 100 ms wide frequency sweep signal from 2 to 40 kHz. Then, we employ fast Fourier transformation on the time-history obtained by each measurement and construct a 2D transmission profile for a frequency slice. For the transient wavefield shown in Fig.~\ref{fig5}, we send a Gauss-modulated sine pulse with 80 cycles at a given center frequency. We apply a %600 Hz wide FIR %
bandpass filter around the center frequencies to eliminate ground noise on the plate. Then, we use a cubic interpolation on the measured grid data for a better visualization of wavefield on the plate.

\section*{Acknowledgments}
%\section*{ACKNOWLEDGMENTS}
We gratefully acknowledge fruitful discussions with Krishanu Roychowdhury (Cornell University), Rui Zhu (Beijing Institute of Technology), and Panayotis Kevrekidis (University of Massachusetts, Amherst). We thank Hiromi Yasuda, Joshua Carter, Minh Nguyen, Sean Lay, and Dzung Tran from the University of Washington for their help in setting up the experiments. We are grateful for the support from NSF (CAREER-1553202 and EFRI-1741685). 
%\end{acknowledgments}

\section*{Author contributions}
C.-W. Chen conceived the design and performed the numerical simulations. R. Chaunsali and C.-W. Chen fabricated the sample. R. Chaunsali performed experiments and interpreted the results. R. Chaunsali and C.-W. Chen wrote the manuscript. J. Yang edited the manuscript and supervised the project.

\end{document}